\documentclass[aps,pre,twocolumn,groupedaddress,showpacs,10]{revtex4-1}
\usepackage{graphicx}
\usepackage{dcolumn}
\usepackage{bm}
\usepackage{epsfig}
\usepackage{epstopdf}
\usepackage{enumerate}
\usepackage{amsmath}

\usepackage{color}
\definecolor{dblue}{rgb}{0,0,0.75}
\definecolor{dred}{rgb}{0.6,0,0}
\definecolor{dgreen}{rgb}{0,0.5,0}
\def\red#1{{\color{black}{#1}\color{black}}}

\newcommand\rout{\bgroup\markoverwith{\textcolor{black}{\rule[0.5ex]{2pt}{0.4pt}}}\ULon}

\newcommand\abs[1]{\left|#1\right|}

\input epsf
\epsfclipon

\begin{document}

\title{Tricriticality in the $q$-neighbor Ising model on  a partially duplex clique}
\author{Anna Chmiel$^{1,2}$, Julian Sienkiewicz$^1$, Katarzyna Sznajd-Weron$^2$} 
\affiliation{
$^1$Faculty of Physics, Warsaw University of Technology, Warsaw, Poland\\
$^2$Department of Theoretical Physics, Wroc{\l}aw University of Science and Technology, Wroc{\l}aw, Poland\\}
\date{\today}
\begin{abstract}
We analyze a modified kinetic Ising model, so called $q$-neighbor Ising model, with Metropolis dynamics, [Phys. Rev. E {\bf 92}, 052105], on a duplex clique and a partially duplex clique. In the $q$-neighbor Ising model each spin interacts only with $q$ spins randomly chosen  from its whole neighborhood. In the case of a duplex clique the change of a spin is allowed only if both levels simultaneously induce this change. Due to the mean-field like nature of the model we are able to derive the analytic form of transition probabilities and solve the corresponding master equation. The existence of the second level changes dramatically the character of the phase  transition. In the case of the monoplex clique, the $q$-neighbor Ising model exhibits continuous phase transition for $q=3$, discontinuous phase transition for $q \ge 4$ and for $q=1$ and $q=2$ the phase transition is not observed. On the other hand, in the case of the duplex clique continuous phase transitions are observed for all values of $q$, even for $q=1$ and $q=2$. Subsequently we introduce a partially duplex clique, parametrized by $r \in [0,1]$, which allows us to tune the network from monoplex ($r=0$) to duplex ($r=1$). Such a generalized topology, in which a fraction $r$ of all nodes appear on both levels, allows to obtain the critical value of $r=r^*(q)$ at which a tricriticality (switch from continuous to discontinuous phase transition) appears. 

\end{abstract}
 \maketitle

\section {Introduction}
Multiplex networks  have become one of the most active area of recent network research \cite{przeg_bianc,przeg_arenas} mainly due to the fact that many real-world systems, like public transport or  social networks, consists of many layers. After seven years of studies we know much more about the structure and  the function of multiplex networks \cite{Battiston:14,Nicosia:15,Dom:15}. A lot of attention has been devoted to the analysis of various dynamics on multiplex networks, including diffusion processes \cite{dyf1}, epidemic spreading \cite{epi_Arenas,epi_plos,prx_moreno} and voter dynamics \cite{marina,ja:15}.

However, there is still an open question how the number of levels influence the macroscopic properties of the  system, such as dynamics of global variables, phase transitions or other emerging patterns. In order to find a general answer to this question systematic studies of various models are needed. However, one of the fundamental problems when dealing with models on the multiplex network is how to ``translate'' the model, originally defined on a monoplex network, into a multiplex network, because usually there are several possibilities. For example, in case of models defined by Hamiltonians \cite{AT} inter-layer interactions can be introduced by the relations between the coupling constants. However, the results of the model can strongly depend on this inter-layer interaction. Jang et al. \cite{AT} thoroughly analyzed the behavior of the Ashkin-Teller model in various types of inter-layer interaction, and presented a rich phase diagram containing three types of phase transitions -- 1st order, 2nd order, and mixed order. For the models defined by the transition probabilities, like the voter model \cite{voter}, the threshold model \cite{Watt} or the $q$-neighbor Ising model \cite{Arek:15} one way of defining interactions between layers is the adoption of so called \texttt{AND} and \texttt{OR} rules. These rules were proposed for the first time by Lee et al. in \cite{Goh2}, where  the generalized threshold cascade model on multiplex networks has been studied. Authors introduced two kinds on nodes: (1) an \texttt{OR} node was activated as soon as a sufficiently large fraction of its neighbors were active in at least one level, (2) an \texttt{AND} node was activated only if in each and every layer a sufficiently large fraction of its neighbors were active. 

The concept of \texttt{AND} and \texttt{OR} rules has been recently adopted in the $q$-voter model with independence \cite{ja:15}, in which the tricriticality is observed on the monoplex network: for $q \le 5$ an order-disorder phase transition is continuous and for $q \ge 6$ it switches to discontinuous \cite{Nyc:Szn:Cis:12}. On a multiplex network consisting of only \texttt{AND} nodes the phase transition switches from continuous to discontinuous for $q^*=4$ if the number of layers $L \le 3$ and for duplex network $q^* = 5$.

Another model, in which tricriticality has been recently observed, is the $q$-neighbor Ising model with Metropolis dynamics \cite{Arek:15,Par:Hoh:17}, a non-equilibrium modification of the kinetic Ising model. The Ising model has always played a very special role in the statistical physics \textcolor{black}{but recently new unexpected behavior, including tricriticality, has been found in one of its non-equilibrium versions \cite{Arek:15,Par:Hoh:17,Jed:Chm:Szn:17}.} For the original Ising model a continuous phase transition is observed for both the regular lattices as well as monoplex complex networks \cite{Pek:01,Agata:02,Dor:Gol:Men:02}. On the other hand, in the case of network of network topology Suchecki and Ho{\l}yst  observed a discontinuous phase transition \cite{Suchecki:09}. Moreover, recently it has been shown that a seemingly small modification of the kinetic Ising model -- in which a randomly chosen spin interacts only with its $q$ neighbors -- leads to the surprising result on monoplex complete graph, i.e. a  switch from a continuous to a discontinuous phase transition at $q=4$ \cite{Arek:15,Par:Hoh:17}. 

In this paper we ask the same question that has been asked previously within the $q$-voter model in \cite{ja:15}, but this time within the $q$-neighbor Ising  model, namely ``How the additional level will influence the type of the phase transition?''. Analogously, as in \cite{ja:15} we focus on a trivial topology, i.e. duplex clique. The experience gained from the $q$-voter model allows us to predict that the switch from a continuous to a discontinuous phase transition will appear for a smaller value of $q$ in case of a duplex clique than for a monoplex one. Such a result would be also expected from the theory of equilibrium phase transitions, if the additional layer could be treated analogously to the additional dimension \cite{ja:15}. However, we will show that our naive prediction fails in case of the $q$-neighbor Ising model and results are exactly opposite. For duplex clique phase transition becomes continuous for all values of $q$. Because for a monoplex network a discontinuous phase transition is observed for $q \ge 4$ and for duplex cliques the phase transition is continuous for an arbitrary value of $q$, we expect that there is an intermediate topology for which a switch from a discontinuous to a continuous phase transition appears. Therefore we investigate the $q$-neighbor Ising model also on a generalized topology of a partially duplex clique, parametrized by $r \in [0,1]$, which allows us to tune the network from monoplex ($r=0$) to duplex ($r=1$). Such a generalized topology, in which a fraction $r$ of all nodes appear on both levels, allows to estimate the critical value of $r=r^*(q)$ at which the switch from continuous to discontinuous phase transition is observed.

\textcolor{black}{The question that naturally appears here is related to the motivation of this work. Certainly investigated topology is interesting from the point of view of social systems, but why to investigate a modified Ising model in this context? There are at least 3 motivations for this work:
\begin{enumerate}
	\item The $q$-neighbor Ising model has shown very intriguing, unexpected behavior already on the complete graph \cite{Arek:15,Par:Hoh:17,Jed:Chm:Szn:17} and we wanted to check what is a role of additional level for this model. We hoped that this would bring us closer to heuristic understanding of the behavior of the model.
	\item We have investigated already the $q$-voter model on duplex clique and based on our studies we have speculated on the role of additional level of a network and on the relation between the type of phase transition and dimensionality of the system \cite{ja:15}. To check universality of our findings we decided to investigate another non-equilibrium model with binary dynamical variables on the same type of network.
	\item Models of opinion dynamics are often based on Ising-spin variables \cite{Nyc:Szn:Cis:12,Nycz:13,Mob:15,Mel:Mob:Zia:17,Sie:Szw:Wer:16,Kru:Szw:Wer:17,Jed:17}.  From this perspective the $q$-neighbor Ising model could be treated as a model of opinion dynamics with two types of social reposes: conformity and independence \cite{Nycz:13}. Such a simple models with binary opinions have found surprisingly many applications, including diffusion of innovation \cite{Mac:etal:16,Byr:etal:16}. Therefore there are particularly worth to be studied on multi-layer networks. 
\end{enumerate}}

\section{Duplex clique  and partially duplex clique}\label{sec:net}
Multiplex networks consist of distinct levels (layers) and the interconnections between levels are only between a node and its counterpart in the other layer (i.e., the same node). {\bf A duplex clique} is a particular case of a multiplex networks, which consists of two distinct levels, each of which is represented by a complete graph (i.e. a clique) of size $N$.  Levels can can be interpreted as two different communities (e.g., Facebook and school class) and are composed of exactly the same people -- each node possesses a counterpart node in the second layer. The same topology was considered to analyze the $q$-voter model with independence \cite{ja:15}. As previously, we assume that each node possesses the same state on each level.

\begin{figure}
\centerline{\epsfig{file=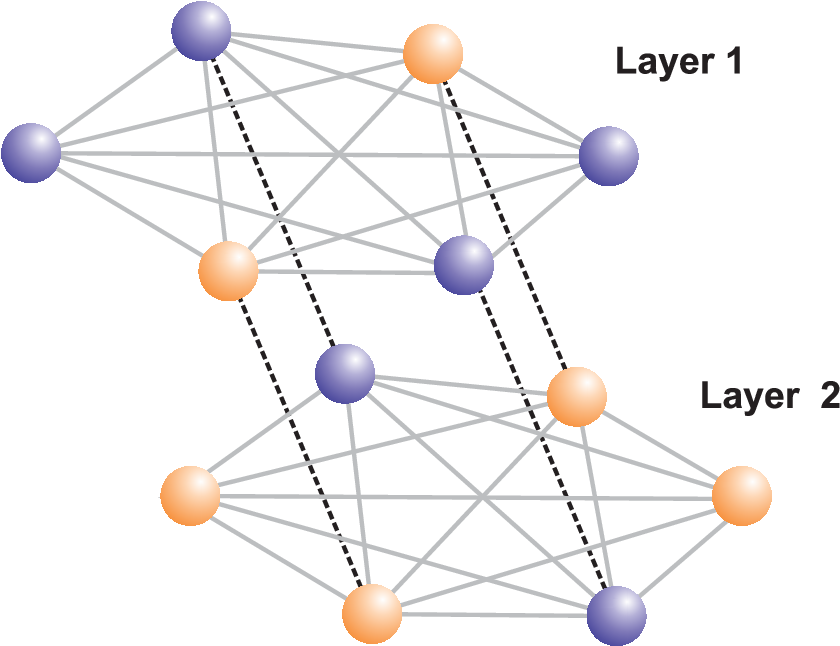,width=.9\columnwidth}}
\caption{(Color online) Example of a partially duplex clique which consists of two complete graphs of size $N=6$ with inter-layer connectivity $r=2/3$.}
\label{fig1}
\end{figure}

In the {\bf partially duplex clique} only fraction $r$ of $N$ nodes have a counterpart in the other layer and all remaining nodes belong only to one community (layer). This means that at each level we have $N_d=Nr$ duplex-type nodes and $N_m=N(1-r)$ monoplex-type nodes, which means that in total the system consists of $N_d+N_m=Nr+2N(1-r)$ distinguishable nodes; an example of such a topology is shown in Fig.~\ref{fig1}. The fraction $r$ of individuals who are active in both layers was   introduced in \cite{epi_plos}. It has been suggested that the parameter $r$ can be interpreted as \emph{interlayer connectivity} or the degree of \emph{structural multiplexity} of the system \cite{marina:16}.

\section{The $q$-neighbor Ising model on duplex clique}\label{Iduplex}

In  \cite{Arek:15} we have modified the kinetic Ising model with Metropolis dynamics allowing each spin to interact only with $q$ spins randomly chosen its neighborhood. We have considered only complete graph, which is particularly convenient for analytical treatment. Here we consider again a set of $N$ spins described by dynamical binary variables $S_i=\pm 1$, but this time they are duplicated on the second level. The algorithm of a single update of the $q$-neighbor Ising model on a duplex clique consists of eight consecutive steps \label{alg}: 
\begin{enumerate}
\item Randomly choose a spin $S_i$
\item From all neighbors of $S_i$ choose a subset $nn_1$ of $q$ neighbors on the first level
\item Calculate the change of the ``energy'' related to the potential flip of spin $S_i$:
\begin{equation}
\Delta E_1 = E_1(-S_i)-E_1(S_i)=2S_i\sum_{j \in nn_1}S_j
\label{eq:qIsing1}
\end{equation}
\item Select randomly a real number $p_1 \in U[0,1]$  and if $p_1< \min[1,e^{-\Delta E_1/T}]$  then set flag $f_1=1$ else $f_1=0$; flag $f_1=1$ indicates a flip, whereas $f_1=0$ suggests to keep the state
\item From all neighbors of $S_i$ choose a subset $nn_2$ of $q$ neighbors on the second level  
\item Calculate the change of the ``energy'' related to the potential flip of spin $S_i$:
\begin{equation}
\Delta E_2 = E_2(-S_i)-E_2(S_i)=2S_i\sum_{j \in nn_2}S_j
\label{eq:qIsing2}
\end{equation}
\item Select randomly a real number $p_2 \in U[0,1]$  and if $p_2< \min[1,e^{-\Delta E_2/T}]$  then set flag $f_2=1$ else $f_2=0$
\item If $f_1=1$ and $f_2=1$ then flip the spin $S_i \rightarrow -S_i$ and its counterpart in the second layer
\end{enumerate}
\begin{figure}
\vskip 0.3cm
\centerline{\epsfig{file=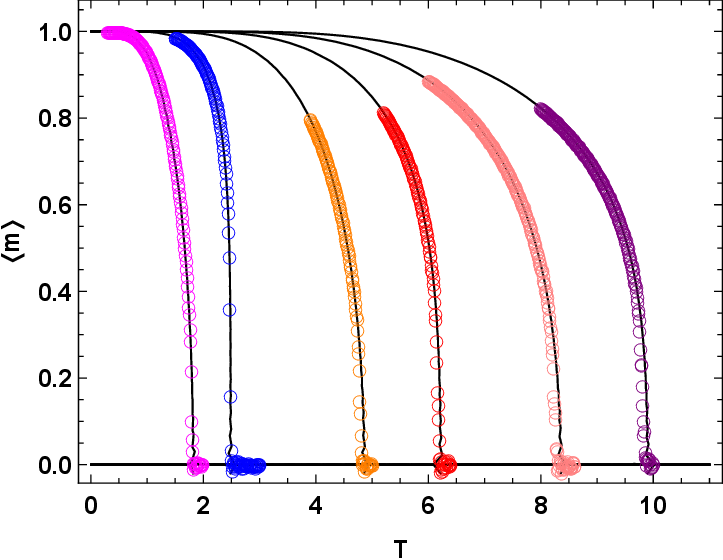,width=.9\columnwidth}}
\caption{(Color online) The average magnetization $\langle m \rangle$ as a function of the temperature $T$ for the duplex clique with $q$ changing from $1$ to $6$ (from left to right). Monte Carlo results (denoted by symbols) were obtained for the system of size $N=10^4$ and averaged over $R=200$ realizations. Lines represent solutions of Eq.~(\ref{F0}).}
\label{fig2}
\end{figure}
We are aware that $E_1$ and $E_2$ are not real energies, and therefore we use quotes, because we sum up interactions with only some randomly chosen neighbors. As usual, a single time step consists of $N$ elementary updates , i.e. $\Delta_t = 1/N$, which means that one time unit  corresponds to the mean update time of a single individual. As an order parameter we choose magnetization, in sociophysics models interpreted often as a public opinion:
\begin{equation}
m=\frac{1}{N} \sum_{i=1}^N S_i.
\label{eq:mag}
\end{equation}

In a single update the number of spins `up' $N^{\uparrow}$ can change according to the following process:
\begin{equation}
N^{\uparrow} (t + \Delta_t) =\left\{
\begin{array}{ll}
N^{\uparrow}(t) + 1 & \mbox{with prob } \gamma^+,\\
N^{\uparrow}(t) - 1 & \mbox{with prob } \gamma^-,\\
N^{\uparrow}(t) & \mbox{with prob } 1-(\gamma^+ + \gamma^-).\\
\end{array}
\right.
\end{equation}
Simultaneously with $N^{\uparrow}$, magnetization $m$ increases/decreases by $2/N$ or remains constant with the above probabilities.

To calculate transition probabilities $\gamma^+$ and $\gamma^-$ for $N \rightarrow \infty$ it is convenient to use the concentration of 'up' spins, which is related to the magnetization by the simple formula:
\begin{equation}
c=\frac{N^{\uparrow}}{N}=\frac{m+1}{2}.
\label{eq:con}
\end{equation}
The transition probabilities as a function of $c$ and model's parameters $T$ and $q$ have the following form:
\begin{eqnarray}
\gamma^+(c,T,q)&=&(1-c) \left[ \sum_{k=0}^{k=q}{ {q \choose k} c^{q-k}(1-c)^{k}  E(q,k) }\right ]^2,\nonumber \\
\gamma^-(c,T,q)&=&c \left[\sum_{k=0}^{k=q}{ {q \choose k}(1-c)^{q-k}c^{k}  E(q,k)}\right ]^2,\nonumber \\
\label{eq:gamma1}
\end{eqnarray}
where
\begin{equation}
E(q,k)=\min \left[ 1,\exp\left( \frac{2(q-2k)}{T}  \right) \right].
\end{equation}
\red{The above equation comes as a direct consqeuence of Metropolis dynamics (i.e., flip probability equal to $\min[1,e^{-\Delta E/T}]$), where the energy change $\Delta E = 2S_i\sum_{j \in nn}S_j$.
As the sum over the spins in the $q$ neighborhood can be expressed as $\sum_{j \in nn}S_j = k S_i - (q-k)S_i$ with $k=0,...,q$ being the number of neighbors which have the same state as spin $S_i$, after short algebra we arrive at $\Delta E =2(2k-q)$.}
For the average values of concentration we can also write the rate equation \cite{Spi:Kra:Red:01}, which has the following form in the rescaled time $t$:
\begin{equation}
\langle c(t+1) \rangle = \langle c(t) \rangle  + \left( \gamma^+(c,T)-\gamma^-(c,T) \right).
\label{eq:rate_duplex}
\end{equation}
In the stationary state $\langle c(t+1) \rangle = \langle c(t) \rangle$, which is equivalent to the condition that the effective force \cite{Nyc:Szn:Cis:12}:
\begin{equation}
F(c,T,q)=\gamma^+(c,T,q)-\gamma^-(c,T,q)=0. 
\label{F0}
\end{equation}

\begin{table}[]
\centering
\begin{tabular}{lcc}
\hline\hline
$q$ & $T_c$ & value\\
\hline
1 & $\frac{2}{\ln 3}$ & 1.82\\
2 & $\frac{4}{\ln 5}$ & 2.49\\
3 & $\frac{6}{-\ln\left[\frac{1}{7}\left(8 + \frac{9(\sqrt{973}-28)^{\frac{1}{3}}}{7^{\frac{2}{3}}} - \frac{27}{(7(\sqrt{973}-28))^{\frac{1}{3}}}\right)\right]}$ & 4.86\\
4 & $\frac{8}{-\ln\left[ \frac{1}{81}\left( 317-20\sqrt{217}\right)\right]}$& 6.22\\
5 & no compact form & 8.35\\
6 & $\frac{12}{-3\ln\left[\frac{1}{13}\left((9534+13\sqrt{537853})^{\frac{1}{3}}-\frac{1}{(9534+13\sqrt{537853})^{\frac{1}{3}}}-18 \right)\right]}$& 9.90\\
\hline\hline
\end{tabular}
\caption{Critical temperature for duplex cliques for first values of $q$.}
\label{tab1}
\end{table}
In Fig.~\ref{fig2} we compare the results obtained from the Monte Carlo simulations and solutions of Eq.~(\ref{F0}) \red{(see Appendix \ref{app:dup} for explicit solutions in case of $q=1$ and $q=2$)}. Both methods give consistent results and continuous phase transition is visible for all values of $q$. In order to find analytically the value of the critical temperature we can use the method proposed in \cite{Nycz:13}, namely we calculate $T$ for which the following condition is fulfilled:
\begin{equation}\label{dF0}
\abs{\frac{\partial F}{\partial c}}_{c=0.5}=0,
\end{equation}
which in our case takes the form of
\begin{equation}\label{dtc}
\sum_{k=0}^{k=q}{q \choose k}\left(2q - 4k - 1\right)E(q,k) = 0.
\end{equation}
For small values of $q$ it is possible to write down explicitly the exact value of $T_c$ (see Table \ref{tab1}), whereas for $q \gg 1$ \red{we observe a linear growth of $T_c$ with $q$}  that can be approximated by $T_c \approx 2q$ \red{(see Appendix \ref{app:dup} for details)}. 

Let us recall here that for the  $q$-neighbor Ising model on a monoplex clique the behavior is much richer: for $q=1 $ and $q=2$ there is no phase transition, for $q=3$ the phase transition is continuous and for $q \ge 4$ discontinuous \cite{Arek:15}. Moreover, for $q>3$ the hysteresis exhibits oscillatory behavior --- expanding for even values of $q$ and shrinking for odd values of $q$. It is worth to stress that the rule used here corresponds to the \texttt{AND} dynamics \cite{ja:15,Goh2} which means that a change of a spin is possible only if both layers indicate the change. In case of the second rule, a so-called \texttt{OR} dynamics, where for the spins's flip indication from only one level is sufficient, there is no phase transition for the $q$-neighbor Ising model on a duplex clique, regardless of the value of $q$. 

\section{The $q$-neighbor Ising model on a partially duplex clique}

As it has been already mentioned in Sec.~\ref{sec:net} in the partially duplex clique only the fraction $r$ of $N$ nodes has a counterpart in the other layer, remaining nodes belong only to one level (see Fig.~\ref{fig1}). Therefore for intermediate topologies, i.e. $r \in (0,1)$ there are two types of nodes at each level: we have $N_d=Nr$ duplex-type nodes (the state is the same on both levels) and $N_m=N(1-r)$ monoplex-type nodes. The algorithm of a single update of the $q$-neighbor Ising model on a partially duplex clique can be described as follows: 
\begin{enumerate}
\item Choose randomly a level; the first or the second with equal probability $1/2$.
\item From $N$ spins on the selected level choose randomly a single spin $S_i$.
\item If the chosen spin belongs to the subset of duplex nodes, the algorithm looks the same as for the duplex clique. 
\item If the spin belongs to the subset of monoplex nodes then: 
\begin{enumerate}
\item From all the neighbors of $S_i$ choose a subset $nn$ of $q$ neighbors (monoplex nodes have neighbors only on one level).
\item Calculate the change of the ``energy'' related to the potential flip of spin $S_i$:
\begin{equation}
\Delta E = E(-S_i)-E(S_i)=2S_i\sum_{j \in nn}S_j
\label{eq:qIsing}
\end{equation} 
\item Flip the $i$-th spin with probability $\min[1,\mathrm{e}^{-\Delta E/T}] $.
\end{enumerate}
\end{enumerate}

We calculate separately concentration of `up'-spins for the monoplex nodes ($c_m$) and duplex nodes ($c_d$):
\begin{eqnarray}
c_{m} & = & \frac{N_{m}^{\uparrow}}{N_{m}} \\
c_{d} & = & \frac{N_{d}^{\uparrow}}{N_{d}}. 
\end{eqnarray}
Since both layers are equivalent, we can restrict our analysis to a single level.

The transition probabilities for the monoplex nodes that describe transitions $c_m \rightarrow c_m \pm 1/N$  are given by:
\begin{eqnarray}
\beta_{m}^+ & = & (1-r)(1-c_m) \times 
\sum_{k=0}^{k=q}{ {q \choose k} c^{q-k}(1-c)^{k}  E(q,k) },\nonumber \\
\beta_{m}^- & = & (1-r) c_m  \times 
\sum_{k=0}^{k=q}{ {q \choose k}(1-c)^{q-k}c^{k}  E(q,k)}\nonumber \\
\end{eqnarray}
and the corresponding rate equation in the rescaled time:
\begin{equation}
\langle c_{m}(t+1) \rangle = \langle c_{m}(t) \rangle + \left( \beta_{m}^+-\beta_{m}^- \right).
\end{equation}

The transition probabilities for the duplex nodes that describe transitions $c_d \rightarrow c_d \pm 1/N$  are given by:
\begin{eqnarray}
\beta_{d} ^+ & = & r(1-c_d)  \left[ \sum_{k=0}^{k=q}{ {q \choose k} c^{q-k}(1-c)^{k}  E(q,k) }\right ]^2\nonumber \\
\beta_{d} ^- & = & r c_d \left[\sum_{k=0}^{k=q}{ {q \choose k}(1-c)^{q-k}c^{k}  E(q,k)}\right ]^2\nonumber \\
\end{eqnarray}
and the corresponding rate equation in the rescaled time:
\begin{equation}
\langle c_{d}(t+1) \rangle = \langle c_{d}(t) \rangle + \left( \beta_{d}^+ -\beta_{d}^- \right).
\end{equation}

Finally the total number of `up' spins in the single layer is equal to $N^{\uparrow}(t)=N_{m}^{\uparrow}+N_{d}^{\uparrow}$ which after dividing by $N$ gives
\begin{equation}
\langle c(t) \rangle = (1-r)\langle c_m(t) \rangle +r\langle c_d(t) \rangle .
\label{rr}
\end{equation}
As in the fully duplex case, the stationary state $\langle c(t+1) \rangle = \langle c(t) \rangle$ is equivalent to the effective force condition
\begin{equation}
F_{q}(c,T,r) = 0,
\label{F0q}
\end{equation}
where the effective force is defined using the following equation
\begin{equation}
\langle c(t+1) \rangle=\langle c(t) \rangle + F_{q}(c,T,r).
\end{equation}

\begin{figure*}[ht]
\centerline{\epsfig{file=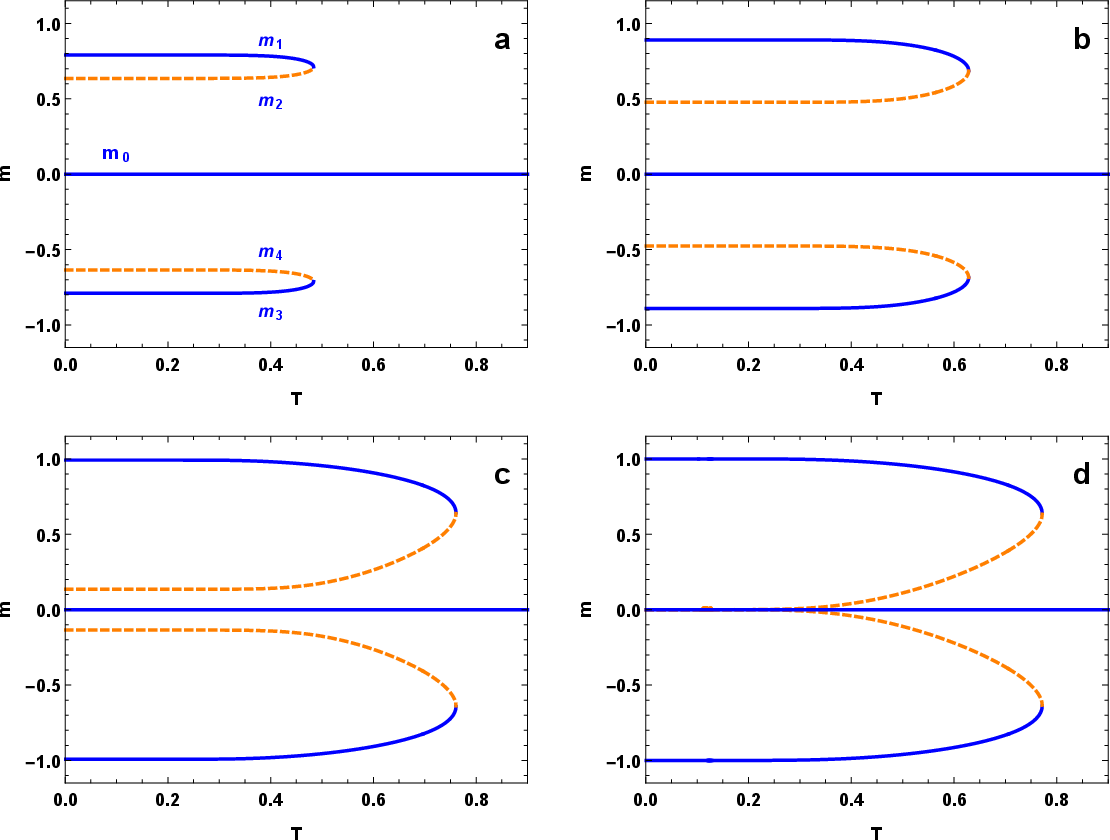,width = .9\textwidth}}
\caption{(Color online) Analytical solutions of magnetization $m$ as a function of temperature  for $q=2$ and (a) $r=0.486$, (b) $r=0.49$, (c) $r=0.499$ and (d) $r=0.5$. Solid lines represent stable solutions while dashed lines show unstable ones.}
\label{fig3}
\end{figure*}

\subsection{Results for $q = 1$ and $q = 2$}
For a monoplex network the phase transition is not present for $q=1$ and $q=2$. We solve explicitly Eq. (\ref{F0q})  for an arbitrary value of  $r$ in case of $q=1$ and $q=2$.  For $q=1$ we obtain the following relation between the critical temperature $T_c$ and $r$ (see Appendix \ref{app:q1} for details):
\begin{equation}
T_c(r)=\frac{2}{\ln\frac{r+2}{r}}.
\label{tc:q1}\end{equation}
Phase transition appears for any $r>0$ and it is continuous (when $r \to 0 $ we have $T_c \to 0$ and no phase transition). For $r=1$ we have $T_c=\frac{2}{\ln 3}$ confirming the result from Table~\ref{tab1}.

\begin{figure}
\centerline{\epsfig{file=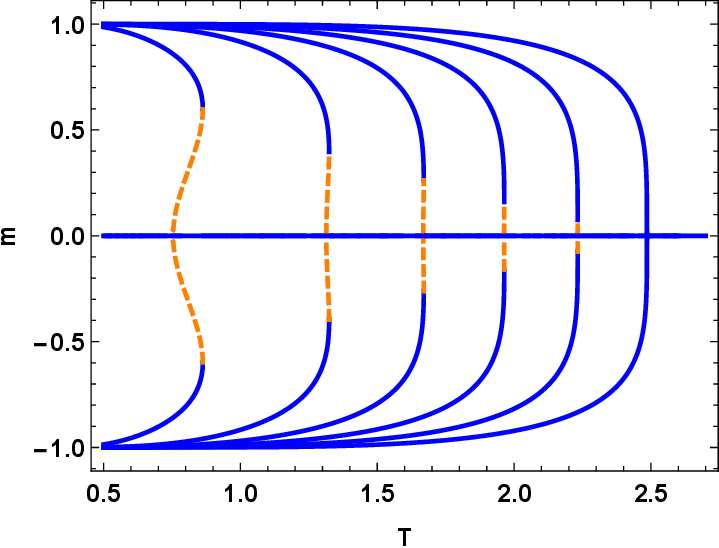,width=.9\columnwidth}}
\caption{(Color online) Analytical solutions for magnetization $m$ as a function of temperature $T$ for $q=2$. Solid lines represent stable (attracting) solutions, dashed lines represent unstable (repellent) solutions: from left to right $r=0.51$, $r=0.6$, $r=0.7$, $r=0.8$, $r=0.9$ and $r=1$. The point where $m=0$ changes its character from an unstable to a stable solution is invisible due to overlapping of lines for different values of $r$.}
\label{fig4}
\end{figure}

In the case of $q=2$ for the narrow range $r$ between $r_c=2(3\sqrt{2}-4)$ and $\frac{1}{2}$ we observe a particularly rich behavior with 5 real solutions, denoted further as $m_0,...,m_4$ (see Fig.~\ref{fig3} and Appendix \ref{app:q1} for explicit equations). The solution corresponding to the disordered state, i.e. $m_0=0$, is stable in the entire range of temperature $T$. Moreover, we have 2 other stable and 2 unstable solutions or rather one pair of stable solutions $m_1,m_3=-m_1$ and one pair of unstable solutions $m_2,m_4=-m_2$. In the language of nonlinear dynamics we have two symmetrical saddle-node bifurcations, where stable solution annihilates with the unstable one \cite{Str:94}. This is interesting phenomena, that cannot be interpreted in terms of classical phase transition. One should notice that a continuous phase transition corresponds to supercritical pitchfork bifurcation: below the bifurcation point there is one stable fixed point $m_0=0$ and above $m_0=0$ becomes unstable, whereas two new stable fixed points appear on either side of the origin symmetrically located at  $m_+, m_-=-m_+$ \cite{Str:94}. On the other hand, a discontinuous phase transition corresponds to subcritical pitchfork bifurcation, which is in a sense inverted supercritical pitchfork bifurcation. The non-zero symmetric fixed points are unstable and exists below the bifurcation point together with a stable fixed point $m_0=0$. Above the bifurcation point two unstable solutions annihilate and $m_0=0$ becomes unstable. As noted by Strogatz in real physical systems, such an explosive instability is usually opposed by stable solutions of higher-order terms, and this is exactly what is observed in case of discontinuous phase transitions. In these cases, the only difference between phase and bifurcation diagrams comes from the definition of an order parameter that takes non-zero value below the critical point, and zero above. Thus the phase diagram could be obtained from bifurcation diagram by reflection $T \rightarrow -T$. The behavior we observe here for $q=2$ reminds in a sense of discontinuous phase transition, because we have hysteresis, i.e., below the transition point we have 3 attractive fixed points and related 3 basins of attractions.  It approaches regular discontinuous phase transition for $r \rightarrow 1/2$. With increasing $r \in (r_c;\frac{1}{2})$ two unstable fixed points approach each other and finally overlap while two stable fixed points move toward $m=\pm1$, what can be seen in Fig.~\ref{fig3}d.

For $r>\frac{1}{2}$ we observe a typical discontinuous phase transition (see Fig. ~\ref{fig4}). The point where two unstable solutions disappear and $m_0=0$ becomes a stable one is clearly seen for $r=0.51$ in Fig.~\ref{fig4}. With increasing $r$ this transition becomes weakly discontinuous --- stable solutions dominate whereas the unstable one is visible only for a very small range of $r$ . Let us underline here that the analytical solutions are fully consistent with the numerical simulations: in Fig. \ref{fig9}d (Appendix \ref{app:fs}) we show the average absolute magnetization $\langle |m| \rangle$ versus temperature $T$ for $r=0.51$ and system size $N=500~000$ compared with respective analytical solutions. The hysteresis as well as the jump of the order parameter are observed in both methods. It is worth to mention here that the observed behavior is strongly dependent on the system size (see Appendix \ref{app:fs} for details).

\begin{figure}[ht]
\epsfig{file=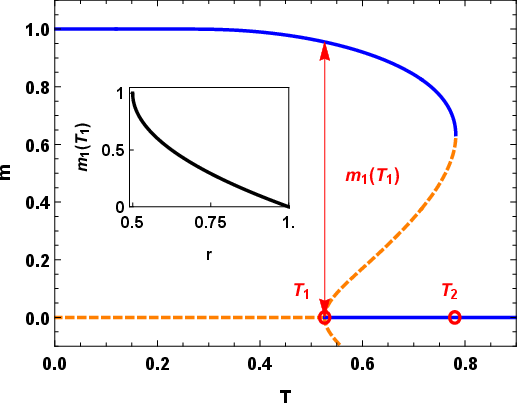,width=.9\columnwidth}
\caption{(Color online) The construction of the $m_1(T_1)$ (red arrow), i.e., the stable solution of an order parameter at the lower spinodal line \red{shown for $r=0.501$}. Inset: the decay of stable magnetization at lower spinodal line with parameter $r$ for $q=2$.}
\label{fig5}
\end{figure}

The natural way to calculate the critical value of parameter $r=r^{*}$, for which the phase transition becomes continuous, would be to look at a distance between upper and lower spinodal lines $T_2-T_1$ (see Fig. \ref{fig5}). While the lower spinodal line $T_1$ can be obtained quite easily from the condition
\begin{equation}\label{Fqc}
\abs{\frac{\partial F_q(c,T,r)}{\partial c}}_{c=\frac{1}{2}}=~0,
\end{equation}
to determine the upper spinodal line $T_2$ is much more tricky. However, we can also obtain $r^{*}$ using other quantity, namely the stable solution of an order parameter at the lower spinodal line $m_1(T_1)$ (or equivalently $m_3(T_1)=-m_1(T_1)$, cf the red arrow in Fig. \ref{fig5}). Please notice that as $T_1$ approaches $T_2$, $m_1(T_1)$ and $m_3(T_1)$ approach zero. Once $m_1(T_1)=m_3(T_1)=0$ the phase transition becomes continuous. The above described procedure is equivalent to solving the equation
\begin{equation}\label{Fq}
F_q(c,T_1(r),r) =  0.
\end{equation}
and obtaining this way $m_1(r) \equiv m_1(T_1(r))$, which can be further used to find the critical value $r^{*}$ by solving $m_1(r)=0$. 

In general for an arbitrary value of $q$ Eq. (\ref{Fq}) is an algebraic equation of high order, however for $q=2$ we have a particularly simple form as in this case $T_1(r) = 4 / \ln\frac{2r+3}{2r-1}$ and $m_1(r) = 1 - \sqrt{2r-1}$. The decay of stable magnetization at lower spinodal line with parameter $r$ is shown in the inset of Fig. \ref{fig5} proving that in case of $q=2$ for all $r > \frac{1}{2}$ we have a discontinuous phase transition as already observed in Fig. \ref{fig4}, \red{while in the limiting case $r \rightarrow 1/2$ we obtain $T_1 = 0$}.

\begin{figure}[ht]
\vskip 0.3cm
\centerline{\epsfig{file=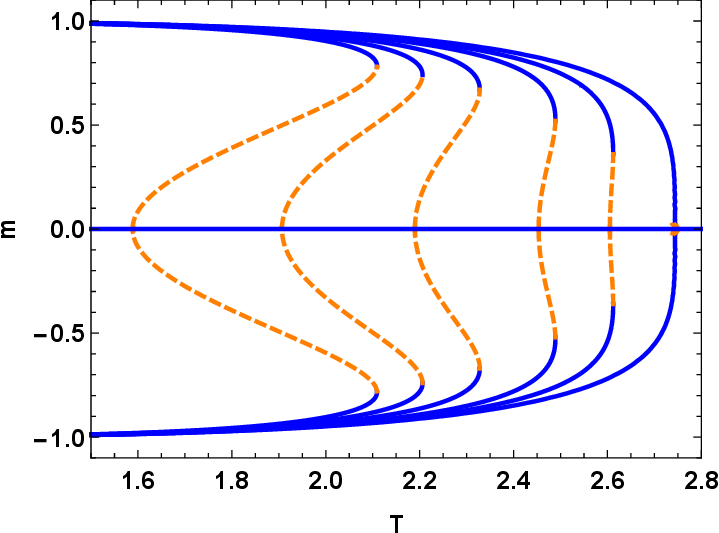,width=.9\columnwidth}}
\caption{Analytical solutions for magnetization $m$ as a function of temperature $T$ for $q=4$ and (from left to right) $r=0$, $r=0.05$,  $r=0.1$, $r=0.15$ $r=0.18$ and $r=0.208$.}
\label{fig6}
\end{figure}

\begin{figure*}
\epsfig{file=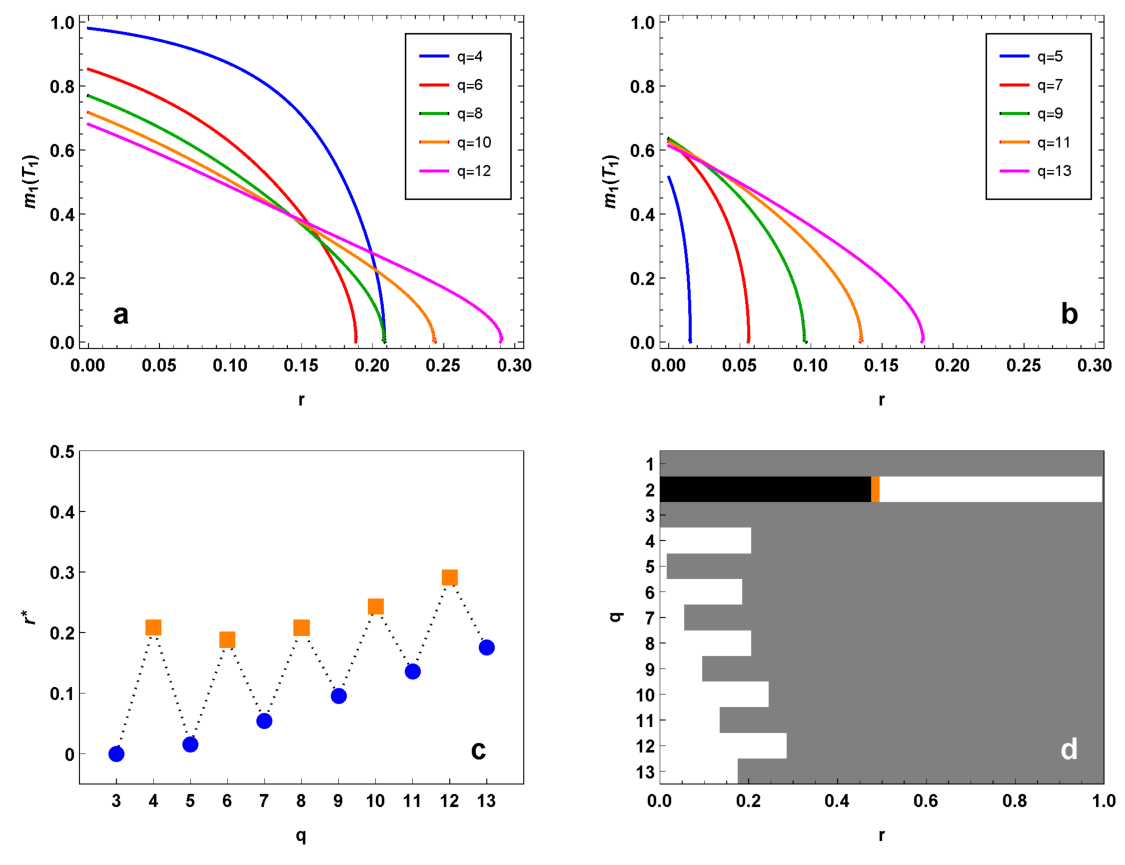,width=.9\textwidth}
\caption{(Color online) (a-b) The value of magnetization $m_1(T_1)$ for even (a) and odd (b) values of $q$. (c) The critical value $r^{*}$ for which the phase transition becomes continuous as a function of $q$. (d) Phase diagram for the $q$-neighbor Ising model on a partially duplex clique in a $(q,r)$-space. White regions indicate discontinuous phase transition and the gray ones --- continuous. The area where phase transition is absent is marked with black. The small range for $q=2$ given with orange represents the special case of phase transition depicted in Fig. \ref{fig3}.}
\label{fig7}
\end{figure*}
  
\subsection{Results for $q>2$}
For $q=3$ the phase transition is continuous for all values of $r$.  This is an expected result since for $q=3$ continuous phase transitions have already been observed for both monoplex \cite{Arek:15} and duplex cliques (see Fig.~\ref{fig2}). On the other hand we have different expectations for $q=4$ where for a monoplex clique the phase transition is discontinuous \cite{Arek:15} whereas for the duplex case it is continuous (Fig.~\ref{fig2}). In Fig.~\ref{fig6} we present magnetization as a function of $T$ for $q=4$ and several values of $r$. It is clearly seen that the unstable regime decreases with the increase of interlayer connectivity $r$. This observation brings us to a natural question that can be formed in the following way ``for which value of $r^*$ the transition becomes continuous?''. In order to cope with that issue we come back to the idea introduced in the previous section for $q=2$ and presented schematically in Fig. \ref{fig5}. The results for $q=4$ as well as for other values up to $q=13$ are shown in Fig.~\ref{fig7}a-b: Fig. \ref{fig7}a presents $m_1(T_1)$ for even values of $q$ whereas Fig. \ref{fig7}b for odd ones. It is seen that for odd values of $q$ function $r^{*}=r^{*}(q)$ is monotonically increasing, while for even values of $q$ the behavior is non-monotonic. This phenomenon is summarized in Fig.~\ref{fig7}c where we also observe  oscillatory behavior of $r^*$ with increasing $q$ for odd and even values of $q$ similar to \cite{Arek:15}.
 
All observations discussed above regarding the character of phase transitions for $q=1,...,13$ are gathered in a $(q,r) - $space phase diagram presented in Fig.~\ref{fig7}d. It shows all three possible outcomes of the model: discontinuous phase transition (marked by white), continuous phase transition (gray), and absence of phase transition (black). The phase diagram underlines apparent differences between even and odd values of $q$. In particular for odd values of $q$ continuous phase transition dominates: for $q=1$ and $q=3$ the transition is continuous for all $r>0$ and for $q=5$ we have discontinuous transition only for a small range of $r \in [0;0.016]$. For larger odd values of $q$ the regime of discontinuity increases. The same situation is observed for even values of $q$ with $q \ge 6$. The two outstanding cases are $q=2$ and $q=4$. One has to stress that $q=2$ is a singular case with a special phase transition occurring for $r \in (2(3\sqrt{2}-4);\frac{1}{2})$ (marked with orange in Fig.~\ref{fig7}d), classic discontinuous phase for $r \in (\frac{1}{2};1)$ and finally continuous phase transition for $r=1$. 

\section{Conclusions}
According to the modern theory of phase transition, each phase transition can be described by an order parameter, having a non-zero value in the ordered phase whereas it vanishes in the disordered phase and this simple classification is used also in the theory of non-equilibrium phase transitions \cite{Hen:Hin:Lue:08}. However, it has been noticed in number of cases that the dichotomy between continuous  and discontinuous  transitions fails, in a sense that jump of the order parameter coincides with power-law singularities \cite{Lip:00,Odo:04,Liu:Sch:Zia:12,Bar:Muk:14,She:Sha:Alv:Koe:Mac:15} or even the absence of hysteresis, metastable states and phase coexistence \cite{Liu:Sch:Zia:12}, which are fundamental indicators of the first-order phase transitions. It can also happen that the phase transition is weakly discontinuous \cite{Fer:etal:92,Sch:Zhe:00}, i.e. the jump of the order parameter is small and therefore to decide if the transition is discontinuous is quite difficult in computer simulations. In such a situation measuring the hysteresis of the order parameter is a demanding task \cite{Odo:04}. However, one should remember that the real size of hysteresis is reached only in thermodynamic limit and for small system it may be unseen. The problem is related to the word ``small'', because for every model the sufficient size of the system in general can be very different. As we have discussed here, for the $q$-neighbor Ising model on a  partially duplex network for size $N=10^4$ the hysteresis is still unseen and even for $N=10^5$ it is still significantly smaller then for the infinite system. Therefore this is so important to have analytical solutions, which in general possible only for mean-field like topologies.  Having results both analytical and obtained from Monte Carlo simulations we can trust that our findings are correct. This is particularly important in this case, because our results are highly unexpected and difficult to understand.

In the first part of this work we have analyzed the  $q$-neighbor Ising model \cite{Arek:15}, on a duplex clique. Adding the second level radically changes the behavior of the model: on a monoplex network model for $q \ge 4$ exhibits discontinuous phase transition, whereas on the duplex network continuous phase transitions are observed for all values of $q$. This is not an obvious result, especially if we recall the findings and argumentations regarding $q$-voter model on multiplex networks \cite{ja:15}. In the $q$-voter model on monoplex network continuous phase transition is observed for $q \in [2,5]$ and for $q \ge 6$ phase transition becomes discontinuous on various monoplex networks\cite{Nyc:Szn:Cis:12,Jed:17}. For duplex network phase transition becomes discontinuous already for $q=5$. Therefore it has been argued that additional level might play a similar role as additional dimension, because in equilibrium statistical mechanics it is common that systems exhibiting a discontinuous phase transition in high space dimensions may display a continuous transition below a certain critical dimension \cite{Hen:Hin:Lue:08}. However, \textcolor{black}{results obtained within the $q$-neighbor Ising model on a duplex clique show that our speculations were wrong. Unfortunately, we are not even able to resolve which of two statements (maybe both) are not universal: (1) additional level of a network play a similar role as an additional dimension, (2) relation between dimensionality and the type of phase transition in non-equilibrium systems reminds the relation observed in equilibrium systems.}

Even more surprising results have been obtained in the second part of work, where partially duplex clique has been introduced. The most intriguing phenomena has been observed for $q=1$ and $q=2$. Let us recall again that for monoplex networks phase transitions appear in the $q$-neighbor Ising model only for $q \le 3$. Because here we use an \texttt{AND} rule to describe interactions with both layers, we have expected that phase transition would appear for $q$ smaller than $3$, but not for $q=1$. We would like to stress here that the continuous phase transition appear in this case for any positive value of $r$, which means that adding even a small fraction of duplex nodes introduces phase transition. However even bigger surprise is the model's behavior for $q=2$. One would expect that if the phase transition appear already for $q=1$ with arbitrary $r>0$, it will be present also for $q=2$ with $r>0$. Yet, for $q=2$ and small values of $r$, model does not exhibit any phase transition. What's happening for larger values of $r$ is probably even more astounding. For $r<r_c = 2 (3 \sqrt{2}-4)$ there is no phase transition and the only \textcolor{black}{steady} state is disordered, i.e. $m=0$. For $r \in [r_c,0.5]$ there are two more ordered \textcolor{black}{steady} states but the phase tradition differs from the regular discontinuous phase transition. Finally, for $r>0.5$ the phase transition becomes classically discontinuous but discontinuity decreases with increasing $r$ up to continuous for $r=1$. 

We would like to focus for a while on the case $q=2$ and $r \in [r_c,0.5]$, for which "exotic" discontinuous phase transition appears. In a sense the phase transition has all properties of classical discontinuous phase transition: order parameter jumps, there is coexistence of states and dependence on the initial conditions (hysteresis). However, if we look at the phase diagram, we immediately see that this transition \textcolor{black}{looks differently than "equilibrium-type" of phase transitions. If we look at this behavior from the perspective of non-linear dynamics we see that for $r>r_c$ we have two symmetrical saddle-node bifurcation points. For increasing $r$ they are approaching each other and for $r>0.5$ a new bifurcation point appears. This is so called subcritical pitchfork bifurcation, which corresponds to "traditional" discontinuous phase transition \cite{Str:94}. Both types of bifurcations are clearly seen in Fig. \ref{fig5}: at $T_1$ there is subcritical pitchfork bifurcation (it is present only for $r>0.5$) and at $T_2$ there is a saddle-node bifurcation. We are aware that the model we consider here is not an equilibrium model but certain dynamical system and therefore it might seem naive to expect "equilibrium-type" of phase transitions here. On the other hand, it is often believed that \cite{Hen:Hin:Lue:08}: \textit{"Much of what is known about equilibrium phase-transitions can be extended to the non-equilibrium case as well"}. Indeed, in most of Ising-type or voter-type non-equilibrium models only traditional phase transitions have been observed \cite{Nyc:Szn:Cis:12,Arek:15,Par:Hoh:17,Jed:17,Jed:Chm:Szn:17,Lip:Gon:Lip:15} and in this context the behavior of the $q$-neighbor Ising model on  a partially duplex clique is particularly rich.} 

For $q \ge 3$ results are not so surprising, especially if one recall the behavior of the model on a single-layer complete graph \cite{Arek:15,Par:Hoh:17}. Because on a duplex clique phase transition is always continuous and on monoplex it is continuous for $q=3$ and discontinuous for $q \le 4$, we expect that for $q \le 4$ there is a critical value of $r=r^*$ below which the transition is discontinuous and above continuous. This is indeed what we have observed. For all $q \le 4$ there is a critical value  of interconnections $r^*=r^*(q)$ above which phase transition becomes continuous. The transition between discontinuous and continuous regime is smooth: jump of the order parameter and hysteresis decreases to zero continuously. Thus at  $r^*=r^*(q)$ we have a tricritical point. It is worth to mention here that tricriticality has been already observed in the $q$-neighbor Ising model three times and each time its origin was slightly different, but in all cases related to some kind of noise  \cite{Par:Hoh:17,Jed:Chm:Szn:17,Mar:Szn:17}. In \cite{Par:Hoh:17} it has been introduced by the additional heat-bath for links, in \cite{Jed:Chm:Szn:17} by the probability $W_0$ of flip in the absence of the energy changes and in \cite{Mar:Szn:17} by rewiring a network at some time intervals $\tau$. 

\textcolor{black}{In the last paragraph of the Introduction we have summarized the motivation for this paper. Our first aim was to understand better the rich behavior of the $q$-neighbor Ising model, observed earlier in \cite{Arek:15,Par:Hoh:17,Jed:Chm:Szn:17}. Although all results obtained so far were not only numerical, but also analytical, we do not feel that we came closer to conceptual understanding of the model. The second aim was to verify the universality of findings from our previous paper \cite{ja:15}. As we have shown here the relation between the number of levels of network and the type of phase transition is not universal. Results obtained for $q$-voter model with noise and  $q$-neighbor Ising model are contradictory.  For many years the dream of statistical physicists has been to develop and establish the theory of non-equilibrium phase transitions. There was a hope, not fully justified, that much of what is known about equilibrium phase-transitions can be extended to the non-equilibrium situation. However, as shown here, this hope not always become reality, and our "equilibrium" intuition maybe completely wrong. }


\section*{Acknowledgements}
This work was supported by funds from the National Science Centre (NCN, Poland) through Grants Nos. 2015/18/E/ST2/00560 and 2014/12/S/ST3/00326 (to A.C.), \red{No. 2015/19/B/ST6/02612 (to J.S.)}, as well as No. 2016/21/B/HS6/01256 (to K.S.-W.). \red{J.S. also acknowledges support in the scope of RENOIR Project by the European Union Horizon 2020 research and innovation programme under the Marie Skłodowska-Curie grant agreement No 691152 and by Ministry of Science and Higher Education (Poland), grant Nos. 34/H2020/2016, 329025/PnH/2016.}

\appendix
\red{
\section{Explicit solutions for the duplex case}\label{app:dup}
In the case of $q=1$ Eq. (\ref{F0}) takes an explicit form of
\begin{widetext}
\begin{equation}
-2~c^3~\mathrm{e}^{-4/T} \left(-1+\mathrm{e}^{2/T}\right)^2+3~c^2~\mathrm{e}^{-4/T} \left(-1+\mathrm{e}^{2/T}\right)^2+c \left(-1-3 \mathrm{e}^{-4/T}+2 \mathrm{e}^{-2/T}\right)+\mathrm{e}^{-4/T} = 0.
\end{equation}
\end{widetext}
Solving the above equation as well as transforming $c$ to $m$ by $m = 2c-1$ leads to
\begin{align}
m_0 ~ ~ &= ~ 0\nonumber\\
m_{1,2} &= \pm ~ \frac{\sqrt{(1+x)(1-3x)}}{1-x}.
\end{align}
with $x = \mathrm{e}^{-\frac{2}{T}}$. In the same manner, for $q=2$ Eq. (\ref{F0}) reads
\begin{figure}
\epsfig{file=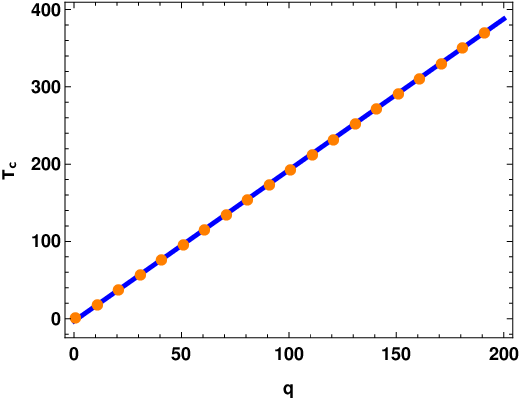,width=.9\columnwidth}
\caption{\red{(Color online) Critical temperature $T_c$ for the duplex clique as a function of $q$. Points represent solutions obtained using Eq. (\ref{dtc}) while solid line comes from a linear regression fit. To prevent visual overlap only every 10-th data point is shown.}}
\label{fig8}
\end{figure}
\begin{widetext}
\begin{equation}
\begin{split}
-2~c^5~\mathrm{e}^{-8/T} \left(-1+\mathrm{e}^{4/T}\right)^2+5~c^4 \mathrm{e}^{-8/T} \left(-1+\mathrm{e}^{4/T}\right)^2+c^3 \left(-6-10 \mathrm{e}^{-8/T}+16 \mathrm{e}^{-4/T}\right)\\
+~c^2 \left(4+10 \mathrm{e}^{-8/T}-14 \mathrm{e}^{-4/T}\right)+c \left(-1-5 \mathrm{e}^{-8/T}+4 \mathrm{e}^{-4/T}\right)+\mathrm{e}^{-8/T} = 0,
\end{split}
\end{equation}
\end{widetext}
which has 5 solutions, with 3 real-valued that can be given as
\begin{align}
m_0 ~ ~ &= ~ 0\nonumber\\
m_{1,2} &= \pm ~ \sqrt{\frac{2\sqrt{(1-5x)(1-x)}-1+5x}{1-x}},
\end{align}
where $x = \mathrm{e}^{-\frac{4}{T}}$. Solutions for higher values of $q$ have increasingly more sophisticated and complex form, nonetheless using Eq. (\ref{dtc}) it is possible to obtain values of critical temperature $T_c$ and plot it as a function of $q$ (see Fig. \ref{fig8}). By applying linear regression we find that $T_c = a q + b$ with $a=1.95474 \pm 0.00067$ and $b=3.093 \pm 0.077$. }

\section{Derivation of analytic equations for $q=1$ and $q=2$ cases with arbitrary $r$}\label{app:q1}

In the case of $q=1$ combining the set of equations (\ref{rr}) with $\beta^{+}_m(c) - \beta^{-}_m(c)=0$ and $\beta^{+}_d(c) - \beta^{-}_d(c)=0$ as well as transforming $c$ to $m$ by $m = 2c-1$ leads to the following three solutions
\red{
\begin{align}
m_0 ~ ~ &= ~ 0\nonumber\\
m_{1,2} &= \pm~ \frac{1+x}{1-x}\sqrt{\frac{(r+2)x - r}{(r-2)x - r}}.
\end{align}}
with $x = \mathrm{e}^{-\frac{2}{T}}$.
\begin{figure*}
\epsfig{file=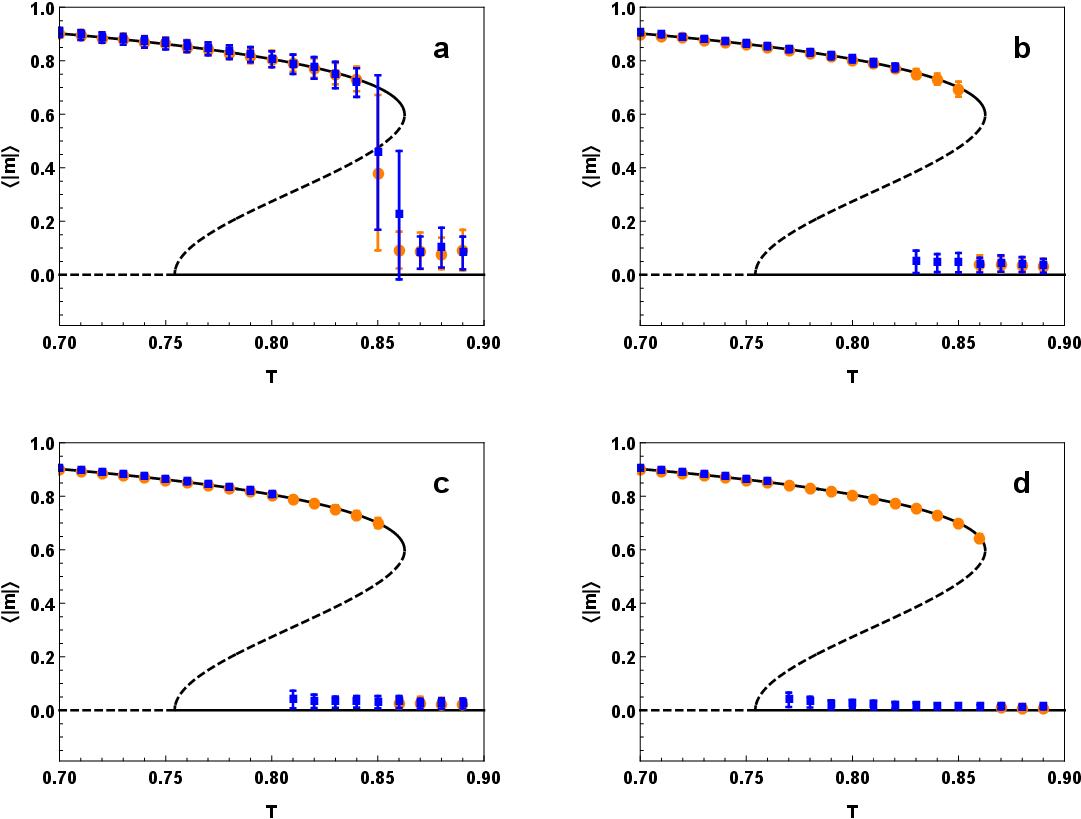,width=.9\textwidth}
\caption{(Color online) Finite-size effects for average absolute magnetization $\langle |m| \rangle$ versus temperature $T$ for $q=2$ and increasing networks size: (a) $N=10~000$, (b) $N=50~000$, (c) $N=100~000$, (d) $N=500~000$. Symbols represent Monte Carlo simulations (squares --- random initial conditions, circles --- ordered initial conditions) while solid and dashed lines give, respectively, stable and unstable solutions obtained from Eqs (\ref{eq:appa2}-\ref{eq:appa4}).}
\label{fig9}
\end{figure*}
The nominator under the radical allows us to get the formula (\ref{tc:q1}) from the main text. A similar procedure performed for $q=2$ results in the following set of five equations:
\red{
\begin{widetext}
\begin{align}\label{eq:appa2}
m_0 ~ ~ &= ~ 0\\
m_{1,3} &= \pm \sqrt{1 - \frac{1}{3}\left[\frac{(x - 1)^2[r^2(x-1)+12r(x-1)+6(x+1)]}{\mathrm{u}(x,r)/(1 - i\sqrt{3})}+\frac{2(x-1)^2[r(x-1) + 6x] + (1+i\sqrt{3})\mathrm{u}(x, r)}{(x - 1)^3}\right]}\\
m_{2,4} &= \pm \sqrt{1 + \frac{2}{3}\left[\frac{(x - 1)^2[r^2(x-1)+12r(x-1)+6(x+1)]}{\mathrm{u}(x,r)}-\frac{(x-1)^2[r(x-1) + 6x] - \mathrm{u}(x, r)}{(x - 1)^3}\right]}
\label{eq:appa4}
\end{align}
\end{widetext}
}
where
\begin{widetext}
\begin{equation}
\begin{split}
\mathrm{u}(x,r) = & \left( 
  3\sqrt{3(x-1)^{15}[4r^3(x-1)^2(r+15+x)-8(x+1)^2(6r(x-1)+x+1)+r^2(-95 + 61 x - 13 x^2 + 47 x^3)]} \right.\\
    & \left. -r (x-1)^8[9+r(18+r)(x-1)+63x] \vphantom{\frac{1}{2}}\right)^\frac{1}{3}
\end{split}
\end{equation}
\end{widetext}
and $x = \mathrm{e}^{-\frac{4}{T}}$. The solutions are shown for exemplary cases in Fig. \ref{fig3}. 

\section{Finite size effects}\label{app:fs}
The importance of the size of the considered system can be examined by comparing the analytical solutions with the Monte-Carlo simulations performed for different number of nodes $N$. Such a setting is shown for $q=2$ in Fig. \ref{fig9}a where we observe a very narrow hysteresis for $N=10~000$ which consequently increases with the system size: $N=50~000$ (Fig. \ref{fig9}b) $N=100~000$ (Fig. \ref{fig9}c) and finally becomes close to analytical solutions ($N=500~000$, Fig. \ref{fig9}d).

\end{document}